\documentclass[twocolumn,pra,showpacs,superscriptaddress]{revtex4}
\usepackage{color}
\usepackage{graphicx}
\usepackage{dcolumn}
\usepackage{bm}
\usepackage{amssymb}
\usepackage{amsmath}
\usepackage{extarrows}

\begin{document}

\title{Generalized Hofstadter model on a cubic optical lattice: From nodal bands to the three-dimensional quantum Hall effect}

\author{Dan-Wei Zhang}
\email{zdanwei@126.com}\affiliation{Guangdong Provincial Key Laboratory of Quantum Engineering and Quantum Materials,
SPTE, South China Normal University, Guangzhou 510006, China}

\author{Rui-Bin Liu}
\affiliation{Department of Physics and Center of Theoretical and Computational Physics, The University of Hong Kong, Pokfulam Road, Hong Kong, China}

\author{Shi-Liang Zhu}
\email{slzhu@nju.edu.cn}
\affiliation{National Laboratory of Solid
State Microstructures and School of Physics, Nanjing University,
Nanjing 210093, China}

\affiliation{Guangdong Provincial Key Laboratory of Quantum
Engineering and Quantum Materials, SPTE, South China Normal
University, Guangzhou 510006, China}

\affiliation{Synergetic Innovation Center
of Quantum Information and Quantum Physics, University of Science
and Technology of China, Hefei, Anhui 230026, China}

\begin{abstract}
We propose that a tunable generalized three-dimensional Hofstadter
Hamiltonian can be realized by engineering the Raman-assisted
hopping of ultracold atoms in a cubic optical lattice. The
Hamiltonian describes a periodic lattice system under artificial
magnetic fluxes in three dimensions. For certain hopping
configurations, the bulk bands can have Weyl points and nodal
loops, respectively, allowing the study of both the two nodal
semimetal states within this system. Furthermore, we illustrate
that with proper rational fluxes and hopping parameters, the
system can exhibit the three-dimensional quantum Hall effect when
the Fermi level lies in the band gaps, which is topologically
characterized by one or two nonzero Chern numbers. Our proposed
optical-lattice system provides a promising platform for exploring
various exotic topological phases in three dimensions.
\end{abstract}

\date{\today}


\maketitle

\section{Introduction}

Topological states of matter have attracted great interest since
the discovery of the quantum Hall effect with intrinsic
topological invariants in two-dimensional electron systems
\cite{IQHE,FQHE}. Early work was devoted to construct theoretical
models in this context \cite{TKNN,Laughlin}, including the
celebrated Haldane model \cite{HaldaneModel} and Hofstadter model
\cite{HHModel}. Following the lines of the two-dimensional work \cite{TKNN},
it has been shown that if there is an energy gap in a three-dimensional periodic lattice, then the integer quantum Hall effect
can result when the Fermi energy lies inside the gap \cite{Halperin,Montambaux1990,Kohmoto1992,Koshino2001}. In the three-dimensional quantum Hall effect, the Hall conductance in each crystal plane can have a quantized Hall value, which is a topological invariant--the first Chern number of a $U(1)$ principal
fiber bundle of a torus spanned by the two quasi-momenta for the crystal plane. However, obtaining the energy spectrum with band gaps for the emergence of quantized Hall conductivities in three-dimensional periodic lattices is a non-trivial task since a motion along the third direction may wash out the gaps of the
perpendicular two-dimensional plane \cite{Montambaux1990,Kohmoto1992,Koshino2001}. In recent years, significant advances have been made in the study of band topology of insulating and semimetallic materials, such as topological
insulators \cite{Kane,Qi} and topological nodal semimetals with
Weyl points or nodal loops (bands crossing along closed
lines instead of isolated points) in three-dimensional momentum space
\cite{Wan,Burkov,SMExp1,SMExp2,NL}. Notably, a three-dimensional Dirac point described by the four-component Dirac equation is composed by two Weyl points with opposite chirality, and they can be separated in momentum space by breaking either time reversal symmetry or inversion symmetry.
The low-energy excitations (Weyl quasi-particles) near the Weyl points are described by the Weyl equation, which is a massless two-component Dirac equation in the chiral representation. These exotic nodal semimetal
states are still rare in real materials or artificial systems \cite{Lu2015,Xiao2015}, and the
three-dimensional quantum Hall effect has predicted or observed
only in systems with extreme anisotropy or unconventional toroidal
magnetic fields
\cite{Koshino2001,Mullen,Balicas,McKernan,Bernevig}.

On the other hand, ultracold atomic gases in optical lattices
provide unparalleled controllability and new avenues to simulate
various quantum states in condensed matter physics
\cite{Lewenstein,ZhangFP}. Recent experimental advances in
engineering artificial gauge field and spin-orbit coupling for
neutral atoms
\cite{Lin1,Lin2,ZhangJ,Cheuk,Liu2016,Dalibard,Galitski,GoldmanRPP,Zhai}
have pushed this system to the forefront for exploring topological
phases
\cite{Zhu2006,Umucalilar,Shao,Goldman2010,Bermudez,Beri,Zhu2011,Sun,Deng,Chen,Zhu2013,Wang,Liu2014}.
For instance, the Zak phase in topologically nontrivial Bloch
bands realized in one-dimensional optical superlattices has been
measured \cite{Bloch2013a}. The Haldane model and the Hofstadter model
have been realized with synthetic magnetic fluxes in
two-dimensional optical lattices
\cite{Jotzu,Bloch2011,Miyake,Bloch2013b,Bloch2015}, and the Chern
number characterizing the topological bands has been measured
\cite{Bloch2015}. The technique of Bloch band tomography in
optical lattices has been proposed and experimentally demonstrated
\cite{BandTomoTheo1,BandTomoTheo2,BandTomoExp1,BandTomoExp2}. In
addition, it has been separately proposed to simulate Weyl points
and nodal loops by engineering atomic Hofstadter bands or
spin-orbit coupling
\cite{Jiang,Dubcek,ZDW2015,He,Xu1,ZDW2016,Xu2,Shastri}. It thus
would be of great value to set an experimentally feasible stage
for quantum simulation of both the two exotic nodal semimetal
states and the rarely explored three-dimensional quantum Hall
effect within an optical-lattice system.

In this paper, we propose an experimental scheme to realize a tunable generalized three-dimensional Hofstadter Hamiltonian with ultracold atoms in a cubic optical lattice based on the Raman-assisted hopping method. The original Hofstadter model describes charged particles moving in a two-dimensional lattice penetrated by a uniform magnetic flux, which develop the quantized energy spectrum and Hall conductivity. The synthetically generated Hamiltonian in our proposal describes a three-dimensional periodic lattice system under three tunable effective magnetic fluxes, which can thus be viewed as a generalized Hofstadter Hamiltonian and be used to simulate various topological phases. We first show that the Weyl points and nodal loops can respectively emerge in the bulk bands of this system for certain hopping configurations, allowing the
study of both the two topological nodal semimetal states within this system. Furthermore, by numerically elaborating the energy spectra and the
topological invariants, we illustrate that for proper rational fluxes and hopping parameters, the system can exhibit the three-dimensional quantum Hall effect when the Fermi level lies in the band gaps, which is topologically characterized by one or two nonzero Chern numbers. Our proposed optical-lattice system provides a powerful platform for exploring exotic topological semimetals and insulators in three dimensions that are rare in real materials.

The paper is organized as follows. Section II
introduces the experimental scheme to realize the tunable three-dimensional Hofstadter Hamiltonian in a cubic optical lattice.
In Sec. III, we show that bulk bands of this system can respectively have Weyl points and nodal loops for certain hopping configurations.
In section IV, we illustrate the emergence of the three-dimensional quantum Hall effect in the system with numerical calculation of the energy spectra and the
topological invariants. Finally, a short conclusion is given in Sec. V.

\section{system and model}

\begin{figure}
\centering
\includegraphics[width=0.45\textwidth]{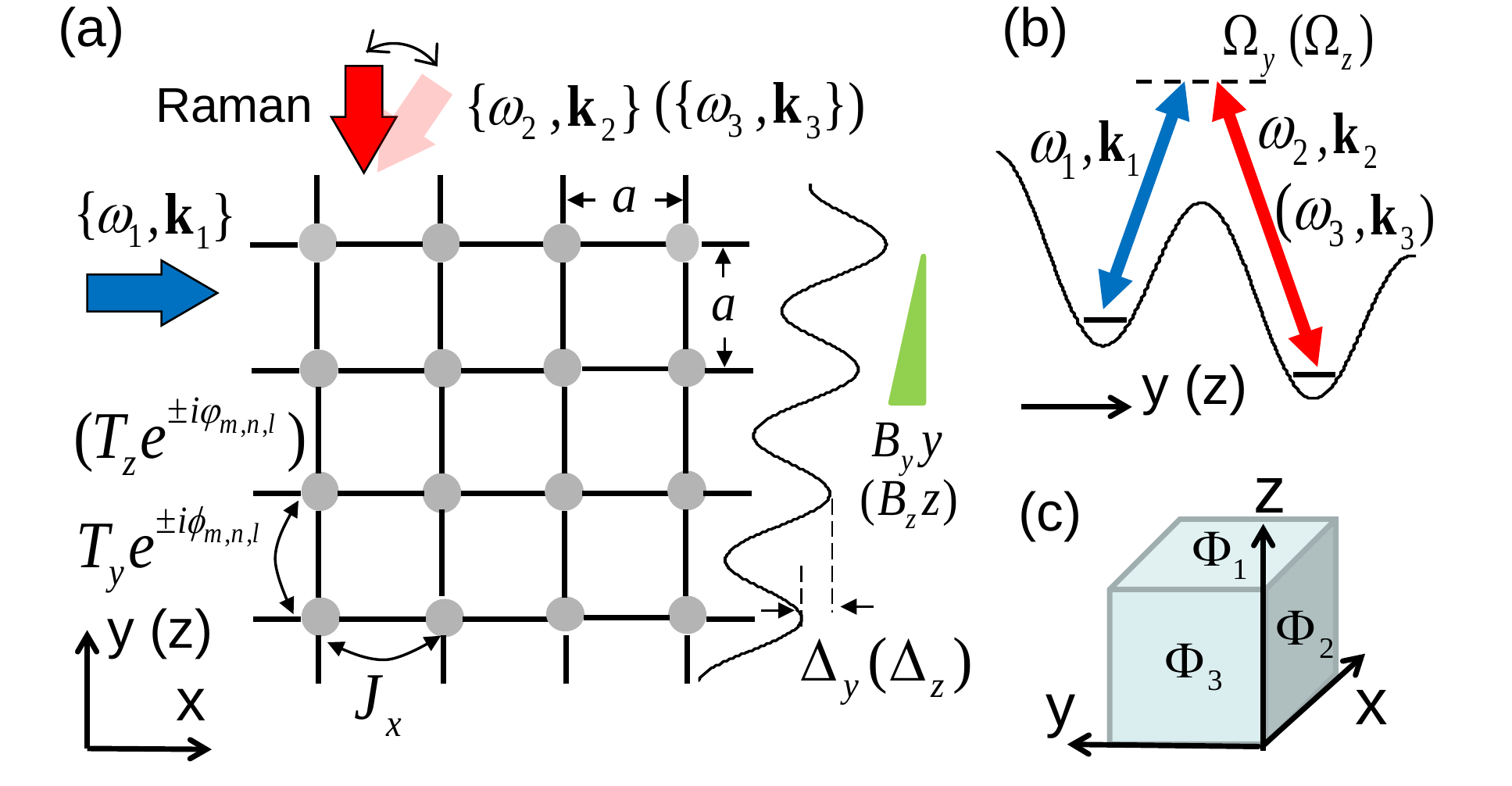}
\caption{(Color online) Proposed realization of a generalized
Hofstadter model in three dimensions. (a) The optical  lattice and
hopping configuration. The atomic hopping along the $x$ axis is
$J_x$. Along the $s$ ($s=y,z$) axis, the tilted lattice with large
tilt potentials $\Delta_{s}$ can be created by the gravity or real
magnetic field gradients $B_ss$. The natural hopping along the $s$
axis is suppressed and then be restored by using three far-detuned
Raman lasers denoted by $\{\omega_j,\mathbf{k}_j\}$ ($j=1,2,3$),
which give rise to complex hopping amplitudes $T_ye^{\pm
i\phi_{m,n,l}}$ and $T_ze^{\pm i\varphi_{m,n,l}}$ with site
indices ($m,n,l$). (b) Laser-assisted tunneling between nearest
neighboring sites along the $s$ axis with the frequency
differences $\omega_2-\omega_1=\Delta_y/\hbar$ and
$\omega_3-\omega_1=\Delta_z/\hbar$ and the effective two-photon
Rabi frequency $\Omega_s$. (c) The effective magnetic fluxes
$\{\Phi_1,\Phi_2,\Phi_3\}$ in the three elementary plaquettes in
the $\{xy,xz,yz\}$ planes, respectively.} \label{system}
\end{figure}

We consider an ultracold degenerate gas of fermionic atoms in a
cubic optical lattice tilted along the $y$ and $z$ axis with $a$
being  the lattice spacing, as shown in Fig. \ref{system}(a). The
atoms are prepared in a hyperfine state of the ground state
manifold, and the tilt potentials with linear energy shift per
lattice site $\Delta_s$ ($s=y,z$) can be generated by the gravity
or real magnetic field gradients $B_ss$, respectively. For the
cases $\Delta_s\gg J_s$ we considered, where $J_s$ denotes the
bare hopping amplitude along the $s$ axis, the atomic hopping
between neighboring sites in these two directions is then
suppressed. To restore and engineer the hopping terms with tunable
effective Peierls phases, we can use the Raman-assisted tunneling
technique \cite{Jaksch,Kolovsky}, which has been experimentally
demonstrated to realize the original Hofstadter model in
two-dimensional optical lattices
\cite{Bloch2011,Miyake,Bloch2013b,Bloch2015}. This method has also
been adopted to simulate Weyl points \cite{Dubcek}. Now we extend
the schemes to realize a generalized three-dimensional Hofstadter
Hamiltonian with fully tunable hopping parameters.

In order to fully and independently engineer the atomic hopping
along the $y$ and $z$ axis in this optical-lattice system, one can
use  three far-detuned Raman beams denoted by their frequencies
and wave vectors $\{\omega_j,\mathbf{k}_j\}$ ($j=1,2,3$), as shown
in Fig. \ref{system}(b). Here we choose the frequency differences
$\omega_2-\omega_1=\Delta_y/\hbar$ and
$\omega_3-\omega_1=\Delta_z/\hbar$ for the resonant tunneling
condition, and assume the effective two-photon Rabi frequencies
$\Omega_s$. Note that we consider $\Delta_y\neq\Delta_z$, which is
required for resonant tunneling along different directions. In
addition, the momentum transfers $\mathbf{Q} = \mathbf{k_1} -
\mathbf{k_2}\equiv(Q_x,Q_y,Q_z)$ and $\mathbf{P} = \mathbf{k_1} -
\mathbf{k_3}\equiv(P_x,P_y,P_z)$ can be independently tunable, for
instance, through independently adjusting the angles of the second
and third Raman lasers with the first Raman laser being fixed, as
shown in Fig. \ref{system}(a). Therefore, the Raman lasers induce
atomic hopping along the $y$ and $z$ axis with tunable spatially
dependent phases $\phi_{m,n,l}=\mathbf{Q}\cdot\mathbf{R}=m\phi_x +
n\phi_y+l\phi_z$ and $\varphi_{m,n,l}=
\mathbf{P}\cdot\mathbf{R}=m\varphi_x+n\varphi_y+l\varphi_z$,
respectively, where $\mathbf{R}=(ma,na,la)$ denotes the position
vector for the lattice site $(m,n,l)$, $\phi_{x,y,z}=aQ_{x,y,z}$
and $\varphi_{x,y,z}=aP_{x,y,z}$. In the high-frequency limit
$\omega_j\gg J_s/\hbar$ \cite{note}, time averaging over rapidly
oscillating Raman beams yields an effective time-independent
Hamiltonian \cite{Miyake,Bloch2013b,Bloch2015}. As a result, the tilts
disappear in the dressed atom picture for resonant tunneling. In
this case, the effective time-independent Hamiltonian for the
three-dimensional lattice system takes the tight-binding form:
\begin{eqnarray} \label{3DHofstadter} \nonumber
\hat{H}&=&-\sum_{m,n,l}J_x\hat a^{\dagger}_{m+1,n,l}\hat a_{m,n,l} + T_ye^{i\phi_{m,n,l}}\hat a^{\dagger}_{m,n+1,l}\hat a_{m,n,l} \\
&&+T_ze^{i\varphi_{m,n,l}}\hat a^{\dagger}_{m,n,l+1}\hat a_{m,n,l} + \text{H.c.},
\end{eqnarray}
where $\hat a^{\dagger}_{m,n,l}$ ($\hat a_{m,n,l}$) is the
creation (annihilation) operator of fermions at the lattice site
$(m,n,l)$, $J_x$ is the natural hopping along the $x$ axis,
$T_ye^{i\phi_{m,n}}$ ($T_ze^{i\varphi_{m,l}}$) denotes the
Raman-induced hopping along the $y$ ($z$) axis with the
spatially-varying phase $\phi_{m,n}$ ($\varphi_{m,l}$) imprinted
by the Raman lasers. The engineered hopping strengths $T_s =
\Omega_s\lambda_s$ can also be tuned through changing the
intensities of the Raman lasers, where $\lambda_s$ denotes the
overlap integral of Wannier-Stark functions between neighbor sites
along the $s$ axis \cite{Miyake,Bloch2013b,Bloch2015}.

The three-dimensional Bravais lattice here is spanned by the
primitive vectors $\mathbf{a}_x$, $\mathbf{a}_y$, and
$\mathbf{a}_z$. One can introduce three effective magnetic fluxes
$\{\Phi_1,\Phi_2,\Phi_3\}$ through the three elementary plaquettes
in the $\{xy,xz,yz\}$ planes with the area $S=a^2$, as shown in
Fig. \ref{system}(c). The effective fluxes, in units of the
magnetic flux quantum, are determined by the phases picked up
anticlockwise around the plaquettes from Hamiltonian
(\ref{3DHofstadter}), which are obtained as
$\Phi_1=\frac{1}{2\pi}(0+\phi_{m+1,n,l}+0-\phi_{m,n,l})=\frac{\phi_x}{2\pi}$,
$\Phi_2=\frac{1}{2\pi}(0+\varphi_{m+1,n,l}+0-\varphi_{m,n,l})=\frac{\varphi_x}{2\pi}$,
$\Phi_3=\frac{1}{2\pi}(\varphi_{m,n,l}+\phi_{m,n,l+1}-\varphi_{m,n+1,l}-\phi_{m,n,l})
=\frac{\phi_z-\varphi_y}{2\pi}$. The three effective magnetic
fluxes can thus be independently tuned from zero  to positive or
negative one, and we here focus on the positive-flux cases since
the opposite flux configurations reproduce the same results.

If one of the atomic hopping strengths along three directions in
Hamiltonian (\ref{3DHofstadter}) becomes zero, the system reduces
to a stack of decoupled two-dimensional Hofstadter systems with
cold atoms \cite{Miyake,Bloch2013b,Bloch2015}. Thus, in general
cases without vanishing hopping, the system corresponds to three
copies of the original Hofstadter model defined in connected
planes. In this sense, Hamiltonian (\ref{3DHofstadter}) can be
regarded as a generalized three-dimensional Hofstadter
Hamiltonian, which describes an effectively charged particle
hopping on a tight-binding cubic lattice in the presence of tunable
uniform magnetic fluxes
\cite{Halperin,Montambaux1990,Kohmoto1992,Koshino2001}.

\begin{figure}
\centering
\includegraphics[width=0.48\textwidth]{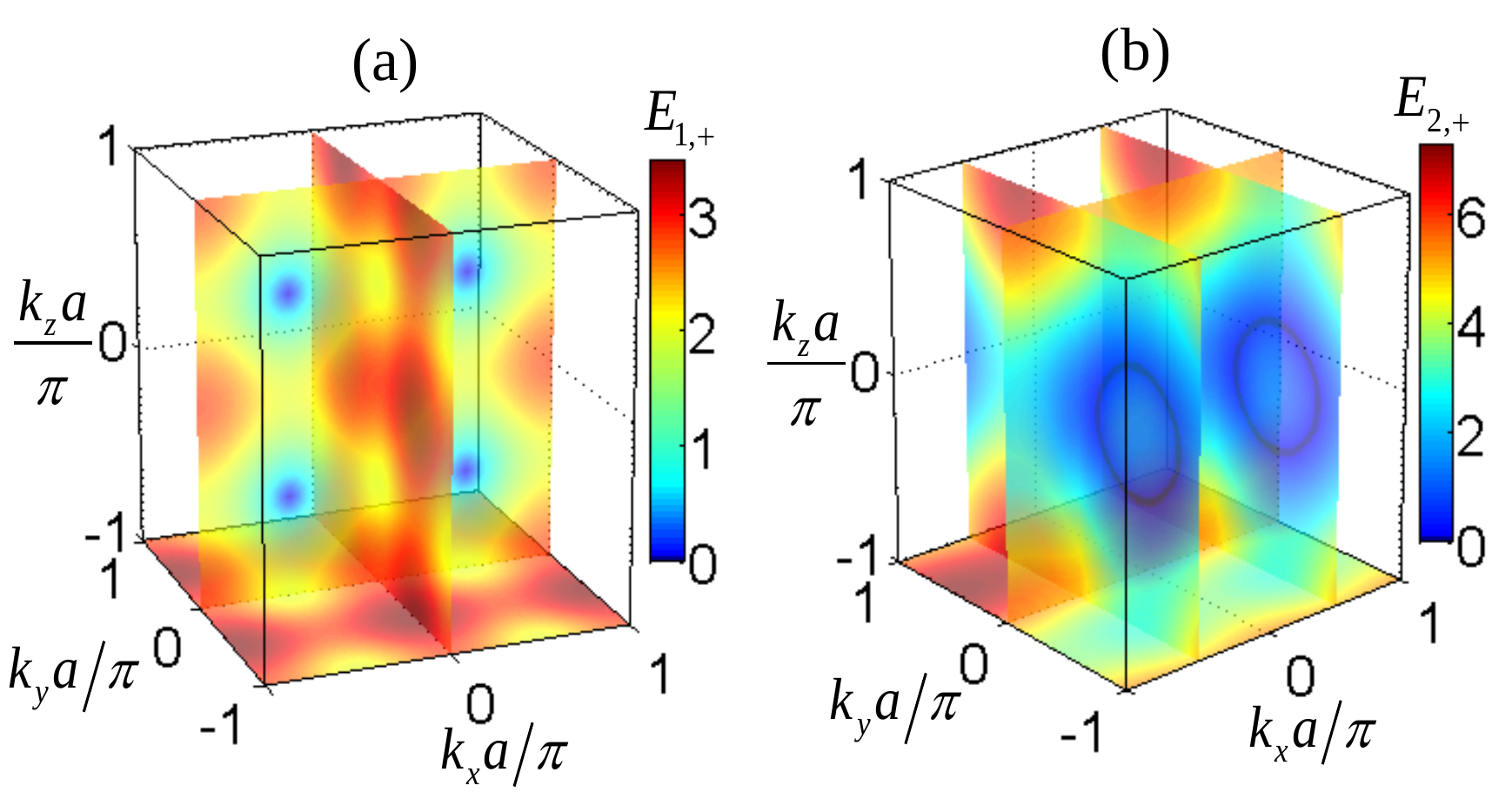}
\caption{(Color online) Simulated Weyl points and nodal loops. The
color-coded three-dimensional bulk spectra:  (a)
$E_{1,+}(\mathbf{k})$ with four Weyl points; and (b)
$E_{2,+}(\mathbf{k})$ with two nodal loops in the $k_x=\pm\pi/2a$
planes. The parameters are $T_y=T_z=\epsilon/3=J_x=1$.}
\label{nodal-point-loop}
\end{figure}

\section{Weyl points and nodal loops}

The hopping and flux parameters in Hamiltonian
(\ref{3DHofstadter}) are tunable through the laser configuration,
enabling the optical-lattice system for simulating various
topological states. In this section, we show that for certain
hopping configurations, the bulk bands of the system can
respectively have Weyl points and nodal loops, which allow one to
study both of the topological Weyl and nodal-loop semimetals
within this cold atom system.

We first choose the hopping parameters
$\phi_{x,y}=\varphi_{x,y}=\phi_z/2=\varphi_z/2=1/2$. In this
configuration,  the hopping phase along the $s$ axis (both the $y$
and $z$ axis) is $\pi$ ($0$), for the site indices $m+n$ odd
(even). Thus the lattice system has two sublattices $A$ and $B$,
which correspond to $m+n$ odd and even, respectively. In this
case, Hamiltonian (\ref{3DHofstadter}) can be rewritten as
\begin{eqnarray} \label{H1}
\hat{H}_1=-\sum_{A}\left( J_x \hat{b}^\dagger_{A+\hat{x}} \hat{a}_{A} - T_y \hat{b}^\dagger_{A+\hat{y}} \hat{a}_{A} - T_z \hat{a}^\dagger_{A+\hat{z}} \hat{a}_{A}\right)\\ \nonumber
-\sum_{B} \left( J_x\hat{a}^\dagger_{B+\hat{x}} \hat{b}_{B} + T_y \hat{a}^\dagger_{B+\hat{y}} \hat{b}_{B} + T_z \hat{b}^\dagger_{B+\hat{z}} \hat{b}_{B}\right)
+ \text{H.c.},
\end{eqnarray}
where $\hat{a}^\dagger_{A}=\hat{a}^\dagger_{B+\hat{\eta}}$ and $\hat{a}_{A}=\hat{a}_{B+\hat{\eta}}$ ($\hat{b}^\dagger_{B}=\hat{b}^\dagger_{A+\hat{\eta}}$ and $\hat{b}_{B}=\hat{b}_{A+\hat{\eta}}$) are the creation and annihilation operators at $A$ ($B$) sublattices with $\eta=x,y,z$ denoting hopping along the $\eta$ direction. The two sublattices form pseudo-spin indices, and Hamiltonian (\ref{H1}) can then be transformed as $\hat{H}_1=\sum_{\mathbf{k}}\hat{\Psi}^{\dag}_{\mathbf{k}}\mathcal{H}_1(\mathbf{k})\hat{\Psi}_{\mathbf{k}}$ with $\hat{\Psi}^{\dag}_{\mathbf{k}}=(\hat{a}^\dagger_{\mathbf{k}},\hat{b}^\dagger_{\mathbf{k}})$ and the Bloch Hamiltonian
\begin{eqnarray} \label{HamWP}
\mathcal{H}_1(\mathbf{k})&=&-2[J_x\cos(k_xa)\sigma_x \\ \nonumber
&&+T_y\sin(k_ya)\sigma_y + T_z\cos(k_za)\sigma_z],
\end{eqnarray}
where $\mathbf{k}=(k_x,k_y,k_z)$ and $\sigma_{x,y,z}$ are the
Pauli matrices acting on the sublattices. The  two bulk bands
$E_{1,\pm}(\mathbf{k})=\pm2\sqrt{J_x^2\cos^2(k_xa)+T_y^2\sin^2(k_ya)+T_z^2\cos^2(k_za)}$
have four Weyl points located at
$\mathbf{k}_W=(\pm\pi/2a,0,\pm\pi/2a)$ in the first Brillouin
zone. Figure \ref{nodal-point-loop}(b) depicts the spectrum
$E_{1,+}(\mathbf{k})$ with the four Weyl points. In the vicinity
of $\mathbf{k}_W$ with $\mathbf{q}=\mathbf{k}-\mathbf{k}_W$, one
can obtain an effective Weyl Hamiltonian,
$\mathcal{H}_{W}(\mathbf{q}) =
v_xq_x\sigma_x+v_yq_y\sigma_y+v_zq_z\sigma_z$, where $v_x=\pm2J_xa$,
$v_y=-2T_ya$, and $v_z=\pm2T_za$ for the four points. The
topological nature of the Weyl points is characterized by their
positive or negative chirality  defined as
$\kappa=\text{sign}(v_xv_yv_z)$. Note that the effective Weyl
Hamiltonian simulated here recovers the one proposed in Ref.
\cite{Dubcek} by a pseudo-spin rotation.

For another hopping configuration given by
$\phi_x=\varphi_x=\phi_{y,z}/2=\varphi_{y,z}/2=1/2$ (here $\phi_z$
and $\varphi_y$ can  alternatively be $\phi_z=\varphi_y=0$), the
nodal loops can exhibit in the bulk bands of the system. In this
case, the atomic hopping along the $y$ axis and the two
sublattices (A and B) are both staggered along the $x$ axis. In
addition, one can add a tunable energy offset between the two
sublattices through a small one-photon detuning
$\epsilon=\Delta_y-\hbar(\omega_2-\omega_1)=\Delta_z-\hbar(\omega_3-\omega_1)$
in the Raman coupling. Under this circumstance, Hamiltonian
(\ref{3DHofstadter}) turns to the form
\begin{eqnarray} \label{H2}\nonumber
\hat{H}_2&=&-\sum_{A}\left( J_x \hat{b}^\dagger_{A+\hat{x}} \hat{a}_{A} - T_y \hat{a}^\dagger_{A+\hat{y}} \hat{a}_{A} - T_z \hat{a}^\dagger_{A+\hat{z}} \hat{a}_{A}\right) \\ \nonumber
&&-\sum_{B}\left( J_x\hat{a}^\dagger_{B+\hat{x}} \hat{b}_{B} + T_y \hat{b}^\dagger_{B+\hat{y}} \hat{b}_{B} + T_z \hat{b}^\dagger_{B+\hat{z}} \hat{b}_{B}\right) \\
&&+\epsilon \sum_{A,B}\left(\hat{a}^\dagger_{A} \hat{a}_{A}-\hat{b}^\dagger_{B} \hat{b}_{B}\right)+\text{H.c.}.
\end{eqnarray}
This gives rise to the Bloch Hamiltonian
\begin{equation} \label{HamWL}
\mathcal{H}_2(\mathbf{k})=-2J_x\cos(k_xa)\sigma_x-g(k_y,k_z)\sigma_z,
\end{equation}
where $g(k_y,k_z)=2T_y\sin(k_ya)+2T_z\cos(k_za)-\epsilon$. The
bulk bands
$E_{2,\pm}(\mathbf{k})=\pm\sqrt{4J_x^2\cos^2(k_xa)+[g(k_y,k_z)]^2}$
have twofold degenerate points located at $k_x=\pm\pi/2a$
planes with $k_y$ and $k_z$ satisfying the condition
$g(k_y,k_z)=0$. The solutions of the condition function can give rise to the
nodal loops for proper parameter $\epsilon$. Figure
\ref{nodal-point-loop}(b) depicts an example of the bulk spectrum
$E_{2,+}(\mathbf{k})$ with two nodal loops in the $k_x=\pm\pi/2a$
planes, respectively.

With fermionic atoms in the optical lattice system, one can
explore the two exotic nodal states, Weyl and nodal-loop
semimetals. In realistic experiments, the simulated nodal points
and loops in the three-dimensional Brillouin zone can be verified
by probing the momentum distribution of the atomic transfer
fraction in the excited band after Bloch-Zener oscillations
through the time-of-flight measurements
\cite{Tarruell,Lim,ZDW2015,ZDW2016}.

\section{The three-dimensional quantum Hall effect}

In this section, we will demonstrate that the proposed cold atom
system can be used to realize the exotic integer quantum Hall
effect in three dimensions. For simplicity but without loss of
generality, we focus on the cases with zero magnetic flux in the
$yz$ plane $\Phi_3=0$, which can be achieved by choosing the
parameters $\phi_z=\varphi_y$. We consider that the effective
magnetic fluxes $\Phi_1$ and $\Phi_2$ are rational, i.e.,
$\Phi_1=p_1/q_1$ and $\Phi_2=p_2/q_2$, with mutually prime
integers $p_{1,2}$ and $q_{1,2}$. In addition, a gauge is chosen
in which the hopping phases $\phi_{m,n,l}=2\pi\Phi_1 m$ and
$\varphi_{m,n,l}=2\pi\Phi_2 m$ in Hamiltonian
(\ref{3DHofstadter}), which corresponds to the Landau gauge in the
continuum case and can be achieved with the parameters, e.g.,
$\phi_y=\varphi_z=2\pi$ and $\phi_z=\varphi_y=0$. The nonvanishing
momentum transfers in the two tilt directions here (i.e.,
$\phi_y=\varphi_z=2\pi$) are necessary for restoring the resonant
tunneling \cite{Miyake,Bloch2013a,Bloch2015}. For the neutral
atomic system within this choice of gauge, the effective vector
potential is given by $\mathbf{A}=(0,B_zx,-B_yx)$, with $\mathbf{B}=(0,B_y,B_z)$ denoting an
effective magnetic field in the $yz$ plane and $x=ma$.

\begin{figure*}
\centering
\includegraphics[width=14cm]{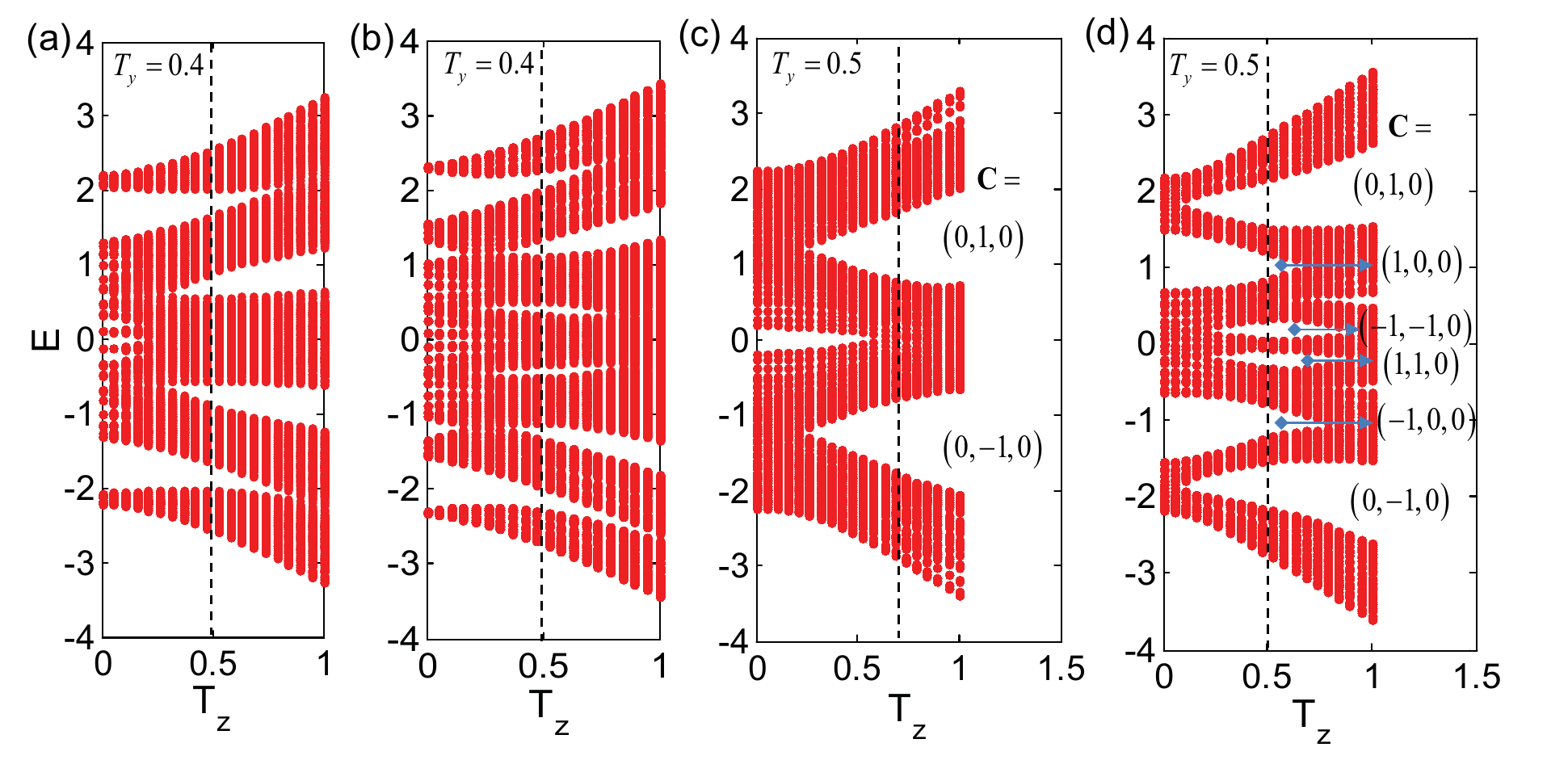}
\caption{(Color online) The energy spectra $E$ as a function of
the hopping strength $T_z$ for typical flux  and hopping
parameters: (a) $\Phi_1=1/5$, $\Phi_2=2/5$, and $T_y=0.4$; (b)
$\Phi_1=1/7$, $\Phi_2=2/7$, and $T_y=0.4$; (c) $\Phi_1=1/2$,
$\Phi_2=1/3$, and $T_y=0.5$; (d) $\Phi_1=1/3$, $\Phi_2=1/5$, and
$T_y=0.5$. In (c) with $T_z=0.7$ and in (d) with $T_z=0.5$ shown
as the dashed lines, the Chern numbers
$\mathbf{C}=(C_{xy},C_{xz},C_{yz})$ when the Fermi level lies in
each energy gap are plotted. The energy unit is set as $J_x=1$.}
\label{Energy1}
\end{figure*}

\subsection{The energy spectra}

An essential point in the quantum Hall effect is that the Fermi
level lies in one of the gaps in the bulk spectrum, such as the
butterfly energy spectrum in the two-dimensional Hofstadter system. So we first study the energy
spectrum in this tunable three-dimensional Hofstadter optical
lattice system. Considering $y=na$ and $z=la$ as the periodic
coordinates on the system, Hamiltonian (\ref{3DHofstadter}) can be
block diagonalized as $\hat{H}=\bigoplus \hat{H}_x(k_y,k_z)$,
where $k_y$ and $k_z$ are the quasimomenta along the periodic
directions and the decoupled block Hamiltonian under the above
conditions takes the following form:
\begin{equation} \label{1DHam}
\hat{H}_x(k_y,k_z)=-\sum_m (J_x\hat a^{\dagger}_{m+1}\hat a_{m} + \text{H.c.}) - \sum_m V_m \hat a^{\dagger}_{m}\hat a_{m},
\end{equation}
where $V_m=2T_y\cos(2\pi\Phi_1m+k_ya)+2T_z\cos(2\pi\Phi_2m+k_za)$.
The corresponding single-particle wave function $\Psi_{mnl}$ is
written as $\Psi_{mnl}=e^{ik_yy+ik_zz}\psi_m$, and then the
Schr\"{o}dinger equation
$\hat{H}_x(k_y,k_z)\Psi_{mnl}=E\Psi_{mnl}$ reduces to a
generalized Harper equation \cite{Harper} with the parameters
$k_y$ and $k_z$ as
\begin{equation} \label{Eigenfun1}
-J_x(\psi_{m-1}+\psi_{m+1})-V_m\psi_m = E \psi_m.
\end{equation}
This reduced one-dimensional tight-binding system with two
commensurabilities $\Phi_1$ and $\Phi_2$ has a period of the least
common multiple of integers $q_1$ and $q_2$ denoted by
$\tilde{q}=[q_1,q_2]$. Under the periodic boundary condition along
the $x$ axis, the wave function $\psi_m$ satisfies
$\psi_m=e^{ik_xx}u_m(\mathbf{k})$ with
$u_m(\mathbf{k})=u_{m+\tilde{q}}(\mathbf{k})$. Therefore in a
general case, the spectrum of the three-dimensional system in the presence of the
effective magnetic fluxes consists of $\tilde{q}$ energy bands and
each band has a reduced (magnetic) Brillouin zone:
$-\pi/\tilde{q}a\leq k_x\leq\pi/\tilde{q}a$, $-\pi/a\leq
k_y\leq\pi/a$, $-\pi/a\leq k_z\leq\pi/a$. In term of the reduced
Bloch wave function $u_m(\mathbf{k})$, Eq. (\ref{Eigenfun1})
becomes
\begin{equation} \label{Eigenfun2}
-J_x(e^{ik_x}u_{m-1}+e^{-ik_x}u_{m+1})-V_m u_m = E(\mathbf{k}) u_m.
\end{equation}
Since $u_m(\mathbf{k})=u_{m+\tilde{q}}(\mathbf{k})$, the problem
of solving the generalized Harper equation (\ref{Eigenfun2})
reduces to solving the eigenvalue equation, $M\Upsilon=E\Upsilon$,
where $\Upsilon=(u_1,...,u_{\tilde{q}})$ is the wave function for
the $\tilde{q}$ bands and $M$ is the $\tilde{q}\times\tilde{q}$
matrix.

\begin{figure}
\centering
\includegraphics[width=8.5cm]{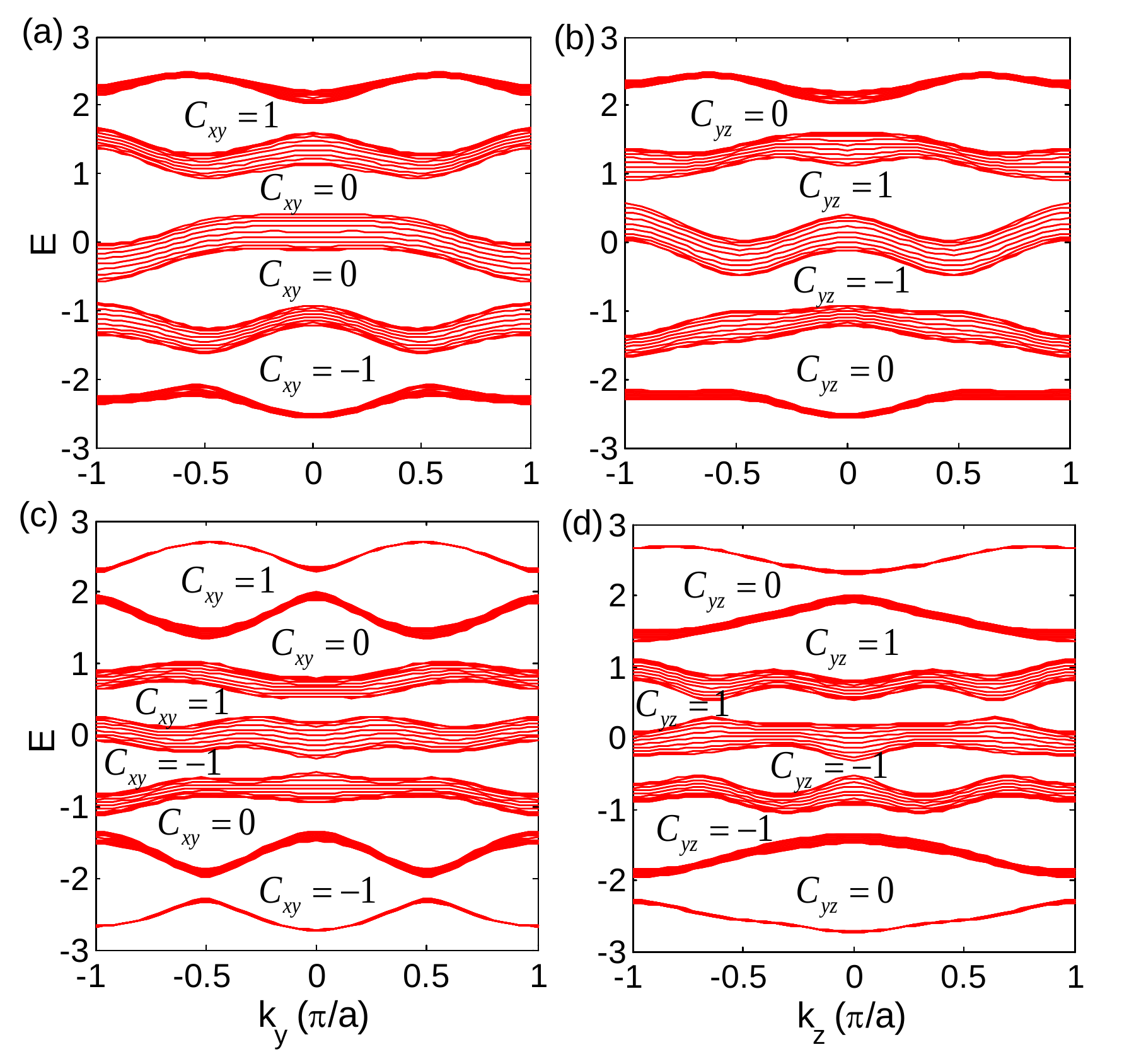}
\caption{(Color online) The energy spectra $E(k_{y})$ and
$E(k_{z})$ and the corresponding Chern numbers $C_{xy}$ and
$C_{xz}$.  (a) and (b) $\Phi_1=1/5$ and $\Phi_2=2/5$; (c) and (d)
$\Phi_1=1/7$ and $\Phi_2=2/7$. We choose $k_z=0$ in (a) and (c),
and $k_y=0$ in (b) and (d). The hopping strengths are $T_y=0.4$
and $T_z=0.5$ in (a-d), as the dashed lines shown in Figs.
\ref{Energy1}(a) and \ref{Energy1}(b). The energy unit is set as
$J_x=1$.} \label{Energy2}
\end{figure}

In order to confirm the above analysis, we numerically calculate the
energy spectra of the system from exact diagonalization of the
tight-binding Hamiltonian (\ref{1DHam}) under the periodic
boundary condition. We first consider the evolution of the spectra
with the increasing of the hopping strength $T_z$ for fixed typical magnetic fluxes and $T_y$
(and $J_x=1$ as the energy unit). As
shown in Fig. \ref{Energy1}, one can find that the overlapping of
the $\tilde{q}$ bands depend on the hopping strength $T_z$.
Figures \ref{Energy1}(a) and \ref{Energy1}(b) show the structures
of these bands for the fluxes $\Phi_2=2\Phi_1=2/5$ with
$\tilde{q}=5$ and $\Phi_2=2\Phi_1=2/7$ with $\tilde{q}=7$,
respectively. In these cases, the largest number of gaps
$N_g=\tilde{q}-1$ is available for the $\tilde{q}$ bands that are
totally non-overlapping, which can be achieved when the hopping
strength $T_z$ is close to $T_y$, such as the dashed lines shown
in Figs. \ref{Energy1}(a) and \ref{Energy1}(b). Figures
\ref{Energy1}(c) and \ref{Energy1}(d) show the bands for the
fluxes $\{\Phi_1=1/2,\Phi_2=1/3\}$ with $\tilde{q}=6$ and
$\{\Phi_1=1/3,\Phi_2=1/5\}$ with $\tilde{q}=15$, respectively. In
these cases, the largest number of gaps always are $N_g<\tilde{q}-1$ for
different $T_z$ because some of the $\tilde{q}$ bands overlap with each
other, such as $N_g=3$ and $N_g=6$ for the dashed lines shown in
Figs. \ref{Energy1}(c) and \ref{Energy1}(d), respectively. For the
cases in Figs. \ref{Energy1}(a) and \ref{Energy1}(b) with the
bands being totally non-overlapping, in Fig. \ref{Energy2}, we
plot the energy spectra for the reduced two-dimensional subsystems
$E(k_y)$ and $E(k_z)$ for fixed $k_z=0$ and $k_y=0$, respectively.
The energy spectra $E(k_y)$ and $E(k_z)$ there reproduce the ones
for the two-dimensional Hofstadter systems with $\tilde{q}=5$
bands in Figs. \ref{Energy2}(a) and \ref{Energy2}(b) and with
$\tilde{q}=7$ bands in Figs. \ref{Energy2}(c) and
\ref{Energy2}(d).

\begin{figure}
\centering
\includegraphics[width=8.5cm]{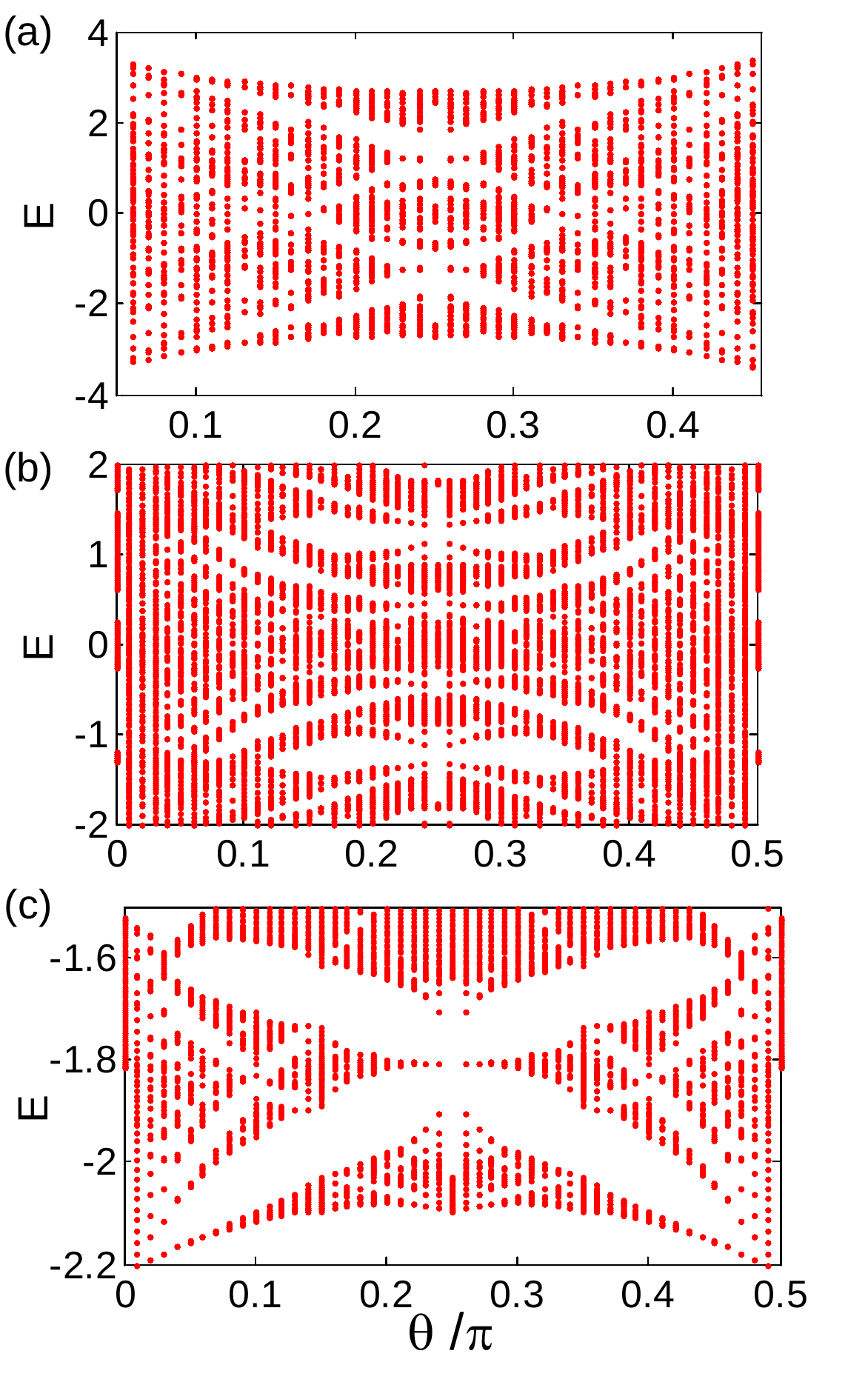}
\caption{(Color online) The energy spectra $E$ as a function of
the angle of the effective magnetic field $\Theta$.  (a)
$T_y=T_z=0.5$ and $B_0=1$; (b) $T_y=T_z=0.5$ and $B_0=0.2$; (b)
$T_y=T_z=0.1$ and $B_0=0.2$. The energy unit is set as $J_x=1$.}
\label{Energy3}
\end{figure}

Due to the fact that the bands overlap in energy or touch with
each other, there is usually no analogous butterfly spectrum with
fractal energy gaps in this three-dimensional generalized
Hofstadter system. It has been shown that the largely anisotropic
hopping (the quasi-one-dimension limit) is required for the
emergence of butterfly-like gaps \cite{Koshino2001}. The results are
confirmed in our numerical calculation under the condition $J_x\gg
T_y,T_z$ with the $x$ axis as the conductive axis. In the
numerical calculation, we have further assumed the effective
magnetic field $\mathbf{B}=(0,B_0\sin\Theta,B_0\cos\Theta)$ with
an angle $\Theta$ tilted from $z$ axis, which corresponds to the
magnetic fluxes $\Phi_3=0$ and
$\Phi_1/\Phi_2=\phi_x/\varphi_x=-\tan\Theta$. In Fig.
\ref{Energy3}, we plot the energy spectra against the angle
$\Theta$ for typical hopping and magnetic strengths. When
$T_y=T_z=0.5J_x$ in Figs. \ref{Energy3}(a) (with $B_0=1$) and
\ref{Energy3}(b) (with $B_0=0.2$), the energy spectra in this
three-dimensional system does not have many gaps (aside from the
trivial Bragg-reflection gaps) with the fractal structure since
many of them are wiped out. In the largely anisotropic case of
$T_y=T_z=0.1J_x$ in Fig. \ref{Energy3}(c), which corresponds to a
quasi-one-dimensional system, a structure akin to the butterfly
seen in the bottom (or at the top) of the whole spectrum emerge
for certain regions of parameters (with $B_0=0.2$). In
experiments, the energy spectrum of this generalized Hofstadter
model may be probed from the density distributions of ultracold
fermions in the optical lattice with an external trap
\cite{LeiWang2014}.

\subsection{The Chern numbers}

It is well-known that the integer quantum Hall effect exhibits in
the two-dimensional Hofstadter system and is topologically
characterized by  a Chern number. Previous work has been devoted
to the generalization of the integer quantized Hall effect in
three-dimensional systems
\cite{Halperin,Montambaux1990,Kohmoto1992,Koshino2001}. It has
been proven that every quantized invariant on a $d$-dimensional
torus $T^d$ is a function of the $d(d-1)/2$ sets of Chern numbers
obtained by slicing $T^d$ by the $d(d-1)/2$ distinct $T^2$
\cite{Avron}. In our three-dimensional Hofstadter system with
$d=3$, the topological invariants for the quantized Hall effect
are given by three Chern numbers
$\mathbf{C}=(C_{xy},C_{xz},C_{yz})$ for three two-dimensional
planes. Here we have $C_{yz}=0$ for the trivial $yz$ plane since
the magnetic flux $\Phi_3$  at this plane is assumed to be zero.
Following the approach in Refs. \cite{Montambaux1990,Kohmoto1992},
when the Fermi energy lies in an energy gap between two bands $N$
and $N+1$ in this system, the other two Chern numbers $C_{xs}$
with $s=y,z$ are given by (let $a=1$)
\begin{equation}\label{ChN}
C_{xs}=\frac{1}{2\pi}\sum_{n\leqslant N}\int_{-\pi}^{\pi}dk_{s'} c_{xs}^{(n)}(k_{s'}),
\end{equation}
where $s'$ denotes replacing $s$ between $y$ and $z$, and
the Chern number $c_{xs}^{(n)}(k_{s'})$ for the $n$-th filling
band (or $n$-th occupied Bloch state) is defined on the torus
$T^2$ spanned by $k_x$ and $k_s$:
\begin{equation}\label{ChN2}
c_{xs}^{(n)}(k_{s'})=\frac{1}{2\pi}\int_{-\pi/\tilde{q}}^{\pi/\tilde{q}}dk_x\int_{-\pi}^{\pi}dk_{s} F_{xs}^{(n)}(\textbf{k}),
\end{equation}
where the corresponding Berry curvature
$F_{xs}^{(n)}(\textbf{k})= \Im(\langle
\partial_{k_{s}}u_{n}(\textbf{k})|\partial_{k_{x}}u_{n}(\textbf{k})\rangle-\langle
\partial_{k_x}u_{n}(\textbf{k})|\partial_{k_{s}}u_{n}(\textbf{k})\rangle)$ is a topological expression as a generalization of the results
in two dimensions \cite{TKNN}.

The determination of the Chern numbers is a delicate procedure
that can be achieved explicitly only in the largely anisotropic cases
with a one-to-one correspondence between the Hall conductivities
on the three-dimensional and two-dimensional Hofstadter systems
\cite{Montambaux1990,Kohmoto1992,Koshino2001}. By generalizing the
efficient way based on the $U(1)$ link to obtain the Berry
curvature in the discrete Brillouin zone \cite{Fukui2005}, we can
numerically calculate the Chern numbers $C_{xs}$ given by Eq.
(\ref{ChN}) for all the regions of hopping parameters. Note that
the Brillouin zone is also discrete in realistic experiments due
to the finite lattice. Here the so-called $U(1)$ link is defined
as $U_{\eta}^{(n)}(\textbf{k}_{\textbf{J}})\equiv\langle
u_{n}(\textbf{k}_{\textbf{J}})|u_{n}(\textbf{k}_{\textbf{J}+\hat{\eta}})\rangle/|\langle
u_{n}(\textbf{k}_{\textbf{J}})|u_{n}(\textbf{k}_{\textbf{J}+\hat{\eta}})\rangle|$
for each pixel $\textbf{k}_{\textbf{J}}=(k_x,k_y,k_z)$ of the
discrete three-dimensional Brillouin zone, where
$\hat{\eta}=\hat{x},\hat{s}$ is a unit vector along the axis. From
the $U(1)$ link, the manifestly gauge-invariant Berry curvature is
given by \cite{Fukui2005}
\begin{equation}
\mathcal{F}_{xs}^{(n)}(\textbf{k}_{\textbf{J}})= i\ln\frac{U_{x}^{(n)}(\textbf{k}_{\textbf{J}})U_{s}^{(n)}(\textbf{k}_{\textbf{J}+\hat{x}})}{U_{x}^{(n)}(\textbf{k}_{\textbf{J}+\hat{s}})U_{s}^{(n)}(\textbf{k}_{\textbf{J}})},\label{U1}
\end{equation}
where $\mathcal{F}_{xs}^{(n)}(\textbf{k}_{\textbf{J}})$ corresponds
to a discrete version of the Berry curvature on the finite lattice.
The Chern numbers $C_{xs}$ can be extracted as
\begin{equation}
C_{xs}=\frac{1}{4\pi^2}\sum_{n\leqslant N}\sum_{\textbf{J}} \mathcal{F}_{xs}^{(n)}(\textbf{k}_{\textbf{J}}).\label{Cn}
\end{equation}
With the above equation, we can numerically calculate the Chern
numbers $C_{xs}$ for the three-dimensional quantum Hall effect
in the absence of the band crossing.

We first consider the cases that all the $\tilde{q}$ bands are
totally non-overlapping with $N_g=\tilde{q}-1$ gaps, as shown  in
Figs. \ref{Energy1}(a,b) and Fig. \ref{Energy2}. In these cases,
the $U(1)$ filed strength $\mathcal{F}_{xs}^{(n)}$ is
well-defined in the whole three-dimensional momentum space as
there is no degeneracy of the Bloch wave functions
$|u_n(\mathbf{k})\rangle$, which are numerically obtained by
solving the eigenvalue equation (\ref{Eigenfun2}). In our
numerical results, we find that $c_{xs}^{(n)}$ are independent
of the momentum $k_{s'}$ in these cases and thus one has a
simpler expression of the Chern numbers $C_{xs}=\sum_{n\leqslant
N}c_{xs}^{(n)}(0)$. In Fig. \ref{Energy2}, the corresponding
Chern numbers $C_{xy}$ and $C_{xz}$ when the Fermi level lies in
each band gap are plotted. In Figs. \ref{Energy2}(a) and
\ref{Energy2}(b), one can find that there is always one nonzero
number $C_{xy}$ or $C_{xz}$ for each of the $N_g=4$ gaps. More
interestingly, as shown in Figs. \ref{Energy2}(c) and
\ref{Energy2}(d), the three-dimensional Hall system can have two
nonzero Chern numbers, such as $C_{xy}=C_{xz}=-1$ and
$C_{xy}=C_{xz}=1$ when the Fermi level lies in the third and
fourth gaps, respectively.

In the presence of band crossing or touching, such as the cases
shown in Fig. \ref{Energy1}(c) and \ref{Energy1}(d), the $U(1)$
Berry curvature is no longer well-defined in the whole momentum space due to certain degeneracies of the
Bloch wave function along the $k_x$ direction. To overcome this
problem in the numerical calculation of the Berry curvature, we
introduce a generalized periodic (twisted) boundary condition
\cite{Niu1985} to Hamiltonian (\ref{1DHam}) with the wave function
of the $n$-th occupied Bloch state:
$|u_n(m+L_x,\alpha,k_y,k_z)\rangle
=e^{i\alpha}|u_n(m,\alpha,k_y,k_z )\rangle$
(here $1\leqslant
m\leqslant L_x$), where $L_x$ denotes the lattice length in the
$x$ axis and $\alpha\in[-\pi,\pi]$ is the twist angle. Under this
boundary condition, the Chern number $c_{xs}^{(n)}(k_{s'})$ in
Eq. (\ref{ChN}) becomes \cite{Niu1985}
\begin{equation}
c_{xs}^{(n)}(k_{s'})=\frac{1}{2\pi}\int_{-\pi}^{\pi}d\alpha\int_{-\pi}^{\pi}dk_{s} F_{xs}^{(n)}(\tilde{\textbf{k}}),
\end{equation}
where $\tilde{\textbf{k}}\equiv(\alpha,k_y,k_z)$ is the
generalized momentum space. By replacing $k_x$
with $\alpha$ [i.e.,
$\textbf{k}_{\textbf{J}}\rightarrow\tilde{\textbf{k}}_{\textbf{J}}=(\alpha,k_y,k_z)$],
we numerically diagonalize the tight-binding Hamiltonian
(\ref{1DHam}) under the twisted boundary condition and obtain the
energies $E_n(\tilde{\textbf{k}})$ and the corresponding
eigenfunctions $|u_n(\tilde{\textbf{k}})\rangle$. The results show
that the energies $E_n(\tilde{\textbf{k}})$ for different $n$ are no
longer overlapped in $\tilde{\textbf{k}}$ space. Therefore, the same procedure of numerically
calculating the Chern numbers $C_{xs}$ based on the $U(1)$-link
method can be implemented by replacing $k_x$ with $\alpha$. As
shown in Figs. \ref{Energy1}(c) and \ref{Energy1}(d), the three
Chern numbers $\mathbf{C}=(C_{xy},C_{xz},C_{yz})$ when the Fermi
level lies in each energy gap are plotted. The results again
demonstrate that the quantum Hall effect in this three-dimensional
Hofstadter system is topologically characterized by one or two
nonzero Chern numbers. We note that the Chern numbers of the Bloch
bands without overlapping in the previous cases can also be
numerically obtained in this way under the twisted boundary
condition, which has been confirmed in our numerical simulations.

In the artificially generated Hofstadter bands in two-dimensional
optical lattices \cite{Miyake,Bloch2013b,Bloch2015}, the Chern
number has been experimentally extracted from the center-of-mass
drift of an ultracold atomic cloud \cite{Bloch2015}. This method
can be directly used in the three-dimensional Hofstadter system to
detect the Chern numbers $C_{xs}$. One can apply a constant
force created by an optical gradient along the $s$ direction, the atomic cloud on the lattice will undergo
Bloch oscillations along this direction. When the filled bands have non-zero Berry curvature $F_{xs}^{(n)}$, the atomic cloud will experience a net drift along the $x$ axis in response to the force and one can measure its center-of-mass evolution to extract the Chern numbers $C_{xs}$ with high accuracy \cite{Bloch2015}, which is analogous to the quantum-Hall-response measurement. The displacement of the atomic cloud along $x$ ($-x$) direction gives the corresponding positive (negative) Chern value. Thus in experiments, for the cases of two nonzero Chern numbers, one can observe two significant displacements of the atomic cloud when varying the direction of the applied force from $y$ axis to $z$ axis, in contract to a single displacement for the cases of one nonzero Chern number. The applied force can be chosen to be strong enough to accurately detect the displacement, but weak enough to limit the Landau-Zener transitions to higher bands. An alternative method to determine the Chern numbers would be directly measuring the Berry
curvature by the newly-developed technique of tomography of Bloch
bands in optical lattices, as proposed and demonstrated in Refs. \cite{BandTomoTheo1,BandTomoTheo2,BandTomoExp1,BandTomoExp2}.

We further consider the energy gap $E_g$ of the topological
nontrivial bands in this system, which is defined by the energy
difference between the top of the filled bands and the bottom of
the empty bands. In our numerical results, we find that the gaps
for filled topological bands with one nonzero Chern number can be large as
$E_g\approx 1$ ($J_x=1$ as the energy unit), such as the filled bands with $C_{xz}=\pm1$ in Fig.
\ref{Energy1} (c). Even for the other bands shown in Fig.
\ref{Energy1}, the corresponding gaps can be up to the hopping
strengths $T_{y,z}$. For the filled bands with two nonzero Chern
numbers shown as the dashed lines in Figs. \ref{Energy1} (b) [also
in Fig. \ref{Energy2}] and \ref{Energy1} (d), the gaps are
numerically obtained as $E_g\approx 0.18$ and $E_g\approx 0.3$,
respectively. So they are also comparable to $T_{y,z}=0.4$ or
$0.5$, which would be large enough for the measurement of their
Chern numbers in realistic experiments
\cite{Bloch2015,BandTomoExp1,BandTomoExp2}. When the Fermi level
lies in the butterfly-like gaps shown in Fig. \ref{Energy3}(c),
there also are one or two nonzero Chern numbers but the gaps there
are small due to the small hopping parameters $T_y=T_z=0.1$. For
instance, when the Fermi energy is $E_F=-0.19$ and
$\Theta=\arctan(3/4)\approx0.2\pi$, the gap for the filled band is
$E_g\approx0.18$, which is one of the primary gaps in the spectrum shown
in Fig. \ref{Energy3}(c). We numerically calculate the Chern
numbers and obtain $\mathbf{C}=(-1,0,0)$ in this case with the fluxes $\Phi_1=3/25$ and $\Phi_2=4/25$. When the
Fermi energy $E_F=-0.19$ and $\Theta=\arctan(5/12)\approx0.126\pi$, the gap for the filled band
in Fig. \ref{Energy3}(c) is $E_g\approx0.05$, and the obtained
Chern numbers are $\mathbf{C}=(1,-1,0)$ with the corresponding
fluxes $\Phi_1=5/65$ and $\Phi_2=12/65$ in this case.

Finally we consider the topological transition between the bands of different Chern numbers $\mathbf{C}$. Without loss of generality, we consider the three-dimensional Hofstader system with parameters and bands shown in Figs. \ref{Energy2}(c) and
\ref{Energy2}(d) and the Fermi energy $E_F$ initially lies in the 3rd gap with two nonzero Chern numbers $\mathbf{C}=(-1,-1,0)$. In experiments, one can observe two significant displacements of the atomic cloud along the $-x$ direction as a Hall response to the applied force from $y$ to $z$ axis. When increasing $E_F$ to lie inside the 4th gap, the topological transition results with the Chern numbers become $\mathbf{C}=(1,1,0)$, such that both the two atomic Hall drifts change direction (along the $x$ direction). Increasing $E_F$ up to be inside the 5th and 6th gaps, two topological transitions happen sequently with the Chern numbers becoming $\mathbf{C}=(0,1,0)$ and $\mathbf{C}=(1,0,0)$, respectively, in which cases one of the two drifts disappears. When all the bands are filled with $\mathbf{C}=(0,0,0)$, there will be vanishing displacement as response to the force of varying directions. For the Fermi energy in the bands, the system is in the conducting phase and the Chern numbers are ill-defined. One can also turn the parameters of the effective magnetic fluxes for fixed $E_F$ to induce the topological transition. For instance, one can vary the parameter $\Theta$ in Fig. \ref{Energy3}(c) from $\pi/8$ to $3\pi/8$ with fixed $E_F=-0.19$. In this case, there are four sequent topological transitions with varying $\mathbf{C}=(1,-1,0)\rightarrow(-1,0,0)\rightarrow(0,-1,0)\rightarrow(-1,1,0)$ for the four primary gaps, which can similarly be revealed from different atomic displacements in the Hall-response measurements.

\section{Conclusion}

In summary, we have proposed an experimental scheme to realize a
tunable generalized three-dimensional Hofstadter Hamiltonian  with
ultracold atoms in a cubic optical lattice, which describes a
lattice system under effective magnetic fluxes in three
dimensions. We have shown that the Weyl points and nodal loops can
respectively emerge in the bulk bands of this system for certain
hopping configurations. Moreover, we have illustrated that with
proper rational fluxes and hopping parameters, the system can
exhibit the three-dimensional quantum Hall effect when the Fermi
level lies in the band gaps, which is topologically characterized
by one or two nonzero Chern numbers. Our proposed optical-lattice
system provides a powerful platform for exploring exotic
topological semimetals and insulators in three dimensions that are
rare in solid-state materials.

\acknowledgements{This work was supported by the NSFC (Grant No.
11604103 and No. 11474153), the NKRDP of China (Grant No.
2016YFA0301803), the NSF of Guangdong Province (Grant No.
2016A030313436), and the Startup Foundation of SCNU.}


\begin{thebibliography}{99}
\bibitem{note} In typical cold-atom experiments, the frequencies of the Raman laser $\omega_j$ is on the order of GHz, and the natual hopping rate $J_s/\hbar$ is less than about 1 kHz.
\end{thebibliography}
\end{document}